\documentclass{scrartcl}
\usepackage{hyperref}

\usepackage{amssymb,amstext,amsmath,amsthm,amsxtra}
\usepackage{color}  
\usepackage{graphicx}

\newcommand{\bea}{\begin{eqnarray}}
\newcommand{\eea}{\end{eqnarray}}
\newcommand{\nn}{\nonumber}
\newcommand{\be}{\begin{equation}}
\newcommand{\ee}{\end{equation}}

\newcommand{\SU}{\mathrm{SU}}
\newcommand{\SO}{\mathrm{SO}}

\newcommand{\nb}{{\bf n}}

\newcommand{\tens}{\otimes}

\newcommand{\dd}{\mathrm{d}}
\newcommand{\R}{\mathbb{R}}

\def\la{\langle}
\def\ra{\rangle}

\theoremstyle{definition}

\theoremstyle{plain}

\begin{document}

\title{On the Expansions in Spin Foam Cosmology}
\author{Frank\ Hellmann
\\Max Planck Institute for Gravitational Physics\\(Albert Einstein Institute)}
\date{\today}

\maketitle

\begin{abstract}
We discuss the expansions used in spin foam cosmology. We point out that already at the one vertex level arbitrarily complicated amplitudes contribute, and  discuss the geometric asymptotics of the five simplest ones. We discuss what type of consistency conditions would be required to control the expansion. We show that the factorisation of the amplitude originally considered is best interpreted in topological terms. We then consider the next higher term in the graph expansion. We demonstrate the tension between the truncation to small graphs and going to the homogeneous sector, and conclude that it is necessary to truncate the dynamics as well.
\end{abstract}

\section{Introduction}
The idea of a spin foam cosmology \cite{Bianchi:2010zs,Vidotto:2010kw} is to take the spin foam approach to quantum gravity seriously and attempt to study the homogeneous sector of the full theory using a series of approximations. The physical meaning of these expansions is of crucial importance for this project. The original paper \cite{Bianchi:2010zs} described three approximations made to the full theory: small graph, single spin foam, and large volume. Then the homogeneous sector within this truncation of the theory was studied. The amplitude associated to the transition from a homogeneous state to another homogeneous state was found to have an intriguing factorisation property. This was interpreted as arising from the factorisation of the Hamiltonian constraint of the classical theory, and taken as evidence for the soundness of the approximations.

More concretely the theory considered in \cite{Bianchi:2010zs} is the KKL formulation \cite{Kaminski2009c} of the EPRL model \cite{Engle:2007wy,Engle:2007qf,Engle:2007uq}. The small boundary graph considered is the graph consisting of two graphs with two vertices and four edges each. Following \cite{Rovelli:2008dx} each of these graphs is called a dipole graph, and can be seen as dual to a degenerate triangulation of $S^3$. This means we are considering spacetimes with boundary $S^3 \cup S^3$. The single spin foam considered was the one obtained by shrinking the full boundary to a single vertex.

The overall aim of this paper is to begin the investigation of the physical meaning of the approximations used. We will sketch conditions the amplitudes have to satisfy in order for low order calculations to be sensible. The idea is simply to compare the behaviour of the next order of the approximation to the behaviour of the previous order. This can be seen as an appropriate implementation of the idea of renormalisation in the current context.

Our main technical point is the derivation of the asymptotic geometry of the five smallest spin foams with dipole boundary containing one vertex, thus giving a geometric picture of the simplest terms of the first order in the vertex expansion. We will also consider what the known asymptotic results for the amplitudes of the larger graph $\Gamma^5$ tell us about the relationship of the graph expansion and the restriction to the homogeneous sector.

There are numerous ambiguities in the precise definition of the model. We will keep the discussion mostly independent of these ambiguities but will point out where our arguments depend on them. A complete assessment of the plausibility of these expansions will crucially depend on these choices though and therefore is beyond the reach of the current paper. Nevertheless an immediate result of our calculations is to reinterpret the factorisation property found in \cite{Bianchi:2010zs} as a topological factorisation. That is, the classical solution seen there lives on the topology $B^4 \cup B^4$.


\section{The vertex expansion}

We begin by considering the vertex expansion. There are two possible views on this expansion, either as the first term in a sum over 2-complexes or as the coarse approximation of a theory defined by a refinement limit. In \cite{Rovelli:2010qx} it was argued that these two possibilities can be the same if the amplitudes satisfy certain conditions. We will assume that the no vertex term represented by the foam $\Gamma^{dipole} \times [0,1]$ corresponds to the identity operator.

\subsection{The one vertex foams}

In this section we will discuss the possible foams with one internal vertex and two dipole graph as boundary. Recall that a generic $n$-complex can be constructed inductively by taking an $n-1$-complex and attaching $n$-balls to it (see e.g. \cite{Hatcher}). In our case we consider 2-complexes with the boundary given by two dipole graphs and one vertex not on the boundary. That means we start with a set of five vertices. We will start by constructing a particularly simple set of five two complexes.

\begin{figure}[htp]
 \centering
 \includegraphics[scale=0.75]{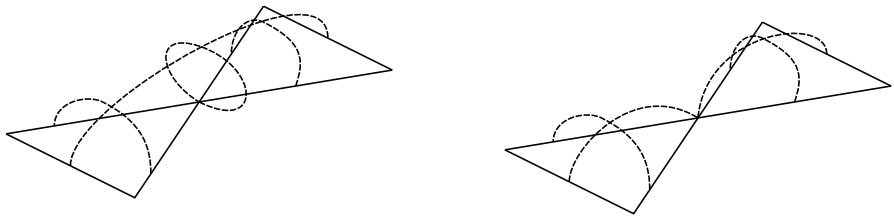}  
 \caption{The two ways to glue discs to the edges. On the left the boundary of the disc is identified with the whole figure eight, on the right two discs are glued to the two triangles individually.}
 \label{fig-gluing}
\end{figure}

We begin by gluing in four edges, connecting the four vertices of the two boundary dipole graphs to the internal vertex. Together with two edges of the boundary graph we have six edges forming a circle with two points identified, or a figure eight. The two simplest ways of gluing faces to these six edges, are by gluing a disc to the whole figure eight, or by gluing two discs to the two circles making up the figure eight, see Figure \ref{fig-gluing}.

\begin{figure}[htp]
 \centering
 \includegraphics[scale=0.75]{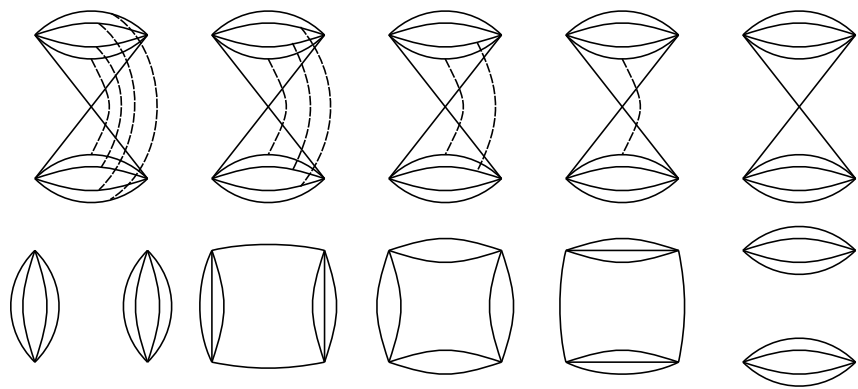}  
 \caption{The one vertex spin foams and their associated KKL spin networks. The dashed lines indicate that the face bounded by the edges it connects does not shrink to a point.}
 \label{fig-TheFoams}
\end{figure}

If we restrict the neighbourhood of the boundary graph $\Gamma$ to look like $[0,1] \times \Gamma$, we can't glue any other edges to the boundary vertices, and only one face to every boundary edge. Thus we obtain five types of vertices, depending on the number of each type of face gluings. These are shown in the top row of Figure \ref{fig-TheFoams}\footnote{The existence of these spin foams was first pointed out by B. Bahr in an unpublished note.}. We can think of the second type of gluing as an edge of the boundary shrinking to a point and then expanding again, the dashed line in Figure \ref{fig-TheFoams} indicates a face that does not shrink to a point, or a gluing with one disc rather than two. In the second row we give the KKL spin networks (see  \cite{Kaminski2009c}) that give the vertex amplitudes of these spin foams. These are simply the intersection of a small sphere around the vertex with the spin foam.

Note that while we will assume that the boundary topology is fixed by the boundary spin network graph to be $S^3 \cup S^3$ we make no such assumption for the bulk topology. While for the 2-complexes that arise, for example, in group field theory and which are dual to triangulations, there always is a natural dual topological space, this is not the case for the more general 2-complexes considered here. Instead we will have to read out the space time topology from the behaviour of the amplitudes. We will argue for the identification of the last foam with the space time topology $B^4 \cup B^4$ in section 2.4, the topology for the other foams is considerably more challenging though.

In \cite{Bianchi:2010zs} only the operator corresponding to the last 2-complex in Figure \ref{fig-TheFoams} was considered. The factorisation property crucial to the physical interpretation in \cite{Bianchi:2010zs} can be seen here already in the fact that in the last KKL spin network there are no edges going from the top to the bottom. This is due to the fact that the spin foam becomes disconnected upon the removal of the vertex, it has the same amplitude as a disconnected foam with two vertices instead of one. In \cite{Bahr:2010my} it was suggested to therefore consider these as equivalent foams, which would place the fifth foam at the second order of the vertex expansion.

Beyond the five simple 2-complexes in Figure \ref{fig-TheFoams} we can construct arbitrarily complicated ones, even subject to the conditions mentioned above. First note that the attaching map for the faces doesn't need to transverse the vertices in the same order, in this way one obtains twisted 2-complexes for which the KKL spin network contain links crossing from the top left corner to the bottom right and vice versa.

Furthermore we can attach more edges to the central vertex, this provides more figure eights, or more generally ``flowers'' with several petals starting at the central vertex. We can then glue further faces to these additional edges. As there is no restriction coming from the topology of the boundary on these, we can glue arbitrarily many faces, creating arbitrarily many links in the KKL spin network. We can distinguish two ways of doing so. If the additional edges and faces become disconnected from the faces and edges we started with upon removal of the central vertex we have a spin foam that factorizes and we speak of a tadpole foam. Its amplitude is a product of a tadpole contribution and the original amplitude. Thus we can divide out the (divergent) sum over the (divergent) amplitudes of all tadpoles.

If, however, we glue the faces touching the boundary to these additional petals these KKL spin networks will be arbitrarily complicated connected networks. In this way we can obtain foams with arbitrarily many faces and arbitrarily complex spin network graphs as amplitudes already at the one vertex level. We will restrict our analysis to the five basic foams depicted above. In particular these are the untwisted foams leading to the smallest KKL spin networks, namely to those with four vertices and eight edges.

\subsection{Asymptotic geometry}

We will now consider the asymptotic geometry of the five basic terms of the first order of the vertex expansion. For completeness we will discuss the asymptotic geometry of the amplitudes for all five terms, including the one discussed in \cite{Bianchi:2010zs}. For simplicity and clarity we will use the standard Perelomov coherent state basis \cite{Livine:2007vk}, rather than the heat kernel coherent states. This will give us a direct interpretation of the geometry in terms of simplicial manifolds. It should be noted that this might not be the only, nor the preferred interpretation of the geometry in this case. We nevertheless expect the salient features of the amplitudes to become clear from these pictures already. As our main interest are the geometric correlations induced by the non-factorising terms we will limit the discussion to the asymptotic geometry of these amplitudes and will not give the full leading order of the large spin expansion of the amplitudes.\footnote{It is worth noting that with the face amplitudes chosen in \cite{Bianchi:2010zs} the first four terms appear to be polynomially suppressed relative to the fifth one, simply by having more faces. However, as described in the last section, there are one vertex spin foams with arbitrarily many faces. These in turn polynomially dominate the fifth term.}

\begin{figure}[htp]
 \centering
 \def\svgwidth{8cm} 
 \includegraphics[scale=1]{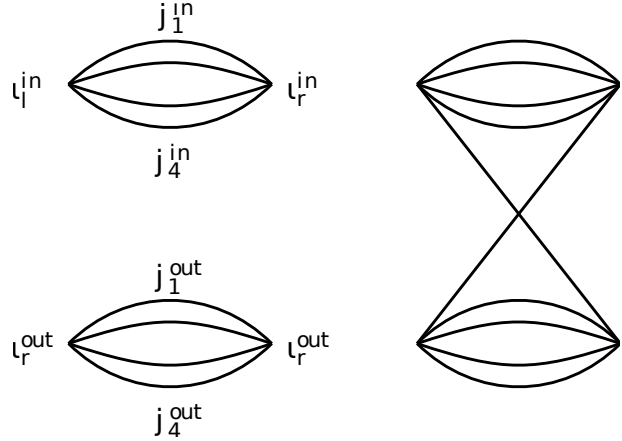}  

 \caption{Our notation for the boundary state space.}
 \label{fig-TheNotation}
\end{figure}

The boundary state space is the spin network space on two dipole graphs. We will arbitrarily label one graph as the $in$ graph and the other as $out$, and one vertex as $l$-eft and the other as $r$-ight. The spins will be denoted $j^{in/out}_i$, with $i = 1\dots4$. We then have four intertwiners $\iota^{in}_l$, $\iota^{in}_r$, $\iota^{out}_l$, $\iota^{out}_r$, living in the invariant subspaces of $\bigotimes_{i=1}^4 j^{in}_i$ and $\bigotimes_{i=1}^4 j^{out}_i$ respectively, see Figure \ref{fig-TheNotation}. They will be labelled by four unit vectors each $\nb^{in}_{li}$, $\nb^{in}_{ri}$, $\nb^{out}_{li}$, $\nb^{out}_{ri}$ each, from which we construct the coherent intertwiners of \cite{Livine:2007vk}. This leaves the phase of the intertwiners unspecified. As we will only discuss the asymptotic geometry it will not enter our discussion here and can be fixed arbitrarily.

The spin network evaluation is given by using the EPRL map $I_{EPRL}$ to boost the $\SU(2)$ intertwiners into $\SU(2)\times\SU(2)$ intertwiners and contracting these using the invariant bilinear inner product $\la,\ra$ according to the pattern given by the KKL spin network\footnote{Note that the analysis would be essentially the same if we just used standard $\SU(2)$ spin networks rather than the EPRL model \cite{Hellmann:2010nf,Barrett:2009as}. This corresponds to setting $\gamma = 1$.}. As the inner product is antisymmetric for half integer representations this only defines the amplitude up to a sign which can be fixed by choosing an orientation on the edges\footnote{We prefer the bilinear inner product as the result only depends on the orientation by sign, not by complex conjugation, and because it implements an orientation preserving gluing in the asymptotic geometry.}. We will only consider the case of Euclidean $\gamma < 1$, writing $\gamma^\pm = \frac{1\pm\gamma}{2}$, so we have \be I_{EPRL}(j_i, \nb_i) = \int_{\SU(2)\times\SU(2)} \dd g^+ \dd g^- \bigotimes_{i=1}^4 g^+ |\nb_i, \gamma^+ j_i\ra \otimes g^- |\nb_i,\gamma^- j_i\ra.\ee

The asymptotic geometry of these boundary intertwiners is given by considering the vectors $j_i \nb_i$ as the outward area normals of a tetrahedron. This is possible as the intertwiners associated to data that does not satisfy closure, $\sum_i j_i \nb_i = 0,$ are exponentially suppressed for large $j$. Thus the picture we have is two unrelated tetrahedra making up the two hemispheres of $S^3$ associated to the left and right vertex of the dipole respectively. The induced geometry is thus discontinuous along the equator of the $S^3$. The two key questions we will investigate are then:

\begin{itemize}
 \item Do the amplitudes correlate the $in$ and $out$ tetrahedra?
 \item Do the amplitudes enforce geometricity between the left and the right hemisphere? That is, do they reduce the discontinuity at the equator?
\end{itemize}

We denote the vertex amplitudes $A_k (j^{in}, \nb^{in}, j^{out}, \nb^{out})$, where $k=0\dots4$ labels the number of edges that contract to a vertex in the spin foam. We will always assume that it is the lowest labelled edges that contract, that is for $A_k$ the edges labelled by $j^{in}_i$ and $j^{out}_i$ with $i \leq k$ contract to a vertex. The other possible amplitudes are obtained by permuting the edges at the boundary graphs.

The first term $A_0$ is simply the propagation by chained EPRL maps acting on the left and right intertwiner separately: \be A_0 = I_{EPRL}^{*} I_{EPRL} \tens I_{EPRL}^{*} I_{EPRL},\ee where $I^*$ denotes the adjoint map under the bilinear inner products. The amplitude of the originally considered graph, $A_4$ is simply the inner product of the EPRL intertwiners on the vertices:

\be A_4 = \left\la I_{EPRL} (j^{in}_l, \nb^{in}_l), I_{EPRL} (j^{in}_r, \nb^{in}_r)\right\ra\left\la I_{EPRL} (j^{out}_l, \nb^{out}_l), I_{EPRL} (j^{out}_r, \nb^{out}_r)\right\ra\ee

This can be written in terms of

\be f(j_i,\nb_i,\nb'_i) = \int_{\SU(2)} \dd g \prod_{i=1}^4 \la - \nb_i|g| \nb'_i\ra^{2j_i},\ee

where $\la|\ra$ is the Hermitian inner product on the fundamental representation. The amplitude then factorises as

\be A_4 = f(\gamma^+ j^{in}_i,\nb^{in}_{li},\nb^{in}_{ri})\;f(\gamma^- j^{in}_i,\nb^{in}_{li},\nb^{in}_{ri})\;f(\gamma^+ j^{out}_i,\nb^{out}_{li},\nb^{out}_{ri})\;f(\gamma^- j^{out}_i,\nb^{out}_{li},\nb^{out}_{ri}).\ee

When evaluated in the coherent intertwiner basis the map $A_0$ can be similarly expressed by switching the roles of $in$/$out$ and $l$/$r$:

\bea A_0 &=& f(\gamma^+ j^{in}_i,\nb^{in}_{li},\nb^{out}_{li})\; f(\gamma^- j^{in}_i,\nb^{in}_{li},\nb^{out}_{li})\times\nn\\&&\times f(\gamma^+ j^{in}_i,\nb^{in}_{ri},\nb^{out}_{ri})\;f(\gamma^- j^{in}_i,\nb^{in}_{ri},\nb^{out}_{ri}) \; \prod_i \delta(j^{in}_i, j^{out}_i) \eea

The asymptotic geometry of $f$ is simply that $f$ is exponentially small unless there is an $\SO(3)$ element $G$ such that $- \nb_{i} = G \nb'_i$, see for example \cite{Livine:2007vk}. If we interpret $j_i \nb_i$ as the outward normals of a tetrahedron this simply says that the two tetrahedra have the same geometry but opposite orientation. From this we can simply read of the asymptotic geometry of the amplitudes $A_0$ and $A_4$. The former simply is that the geometry on the left and right tetrahedron of the $in$ state propagate unperturbed to $out$ state without interacting. The asymptotic geometry of $A_4$ on the other hand has the interpretation that the $in$ and $out$ state are completely uncorrelated while the geometry induced on $S^3$ by the tetrahedra becomes continuous. The left and right tetrahedra can now really be seen as the images of continuous, isometric, orientation preserving maps from the left and right hemisphere of a singularly triangulated, metric $S^3$ into $\R^3$.

We now turn to the amplitude $A_1$ associated to the spin foam where one edge contracts. The KKL spin network for this spin foam is two connected. It is worth noting that in the Lorentzian case the two connected spin networks are naively divergent. As all three or more connected ones were shown to be finite in \cite{Kaminski:2010qb}, the two connected ones would dominate the amplitude. In the Euclidean case which we are studying, we can use Schur's lemma to turn it into the network for $A_0$ rescaled by the inverse dimension of the edge that contracts. It thus has the same geometric interpretation as $A_0$. Writing $d_{j}$ for the signed dimension of the $\SU(2)\times \SU(2)$ irrep $(\gamma^+ j, \gamma^- j)$, we thus have

\be A_1 = \frac{1}{d_{j_1}}A_0.\ee

The amplitude for the case where three edges contract is also two connected and can be rewritten as the amplitude $A_4$, rescaled by the dimension of the edge that does not shrink. Geometrically it thus has the same interpretation again except that the $in$ and $out$ $S^3$  are no longer independent but need to have a face with the same area:

\be A_3 = \frac{1}{d_{j^{in}_4}}\, \delta(j^{in}_4, j^{out}_4)\, A_4.\ee

This leaves the amplitude $A_2$, where two edges contract, as the only case that can not be reduced to $f$. The amplitude is given by:

\bea
A_2 &=& \prod_{\epsilon = \pm} \int_{(\SU(2))^4} \prod_{l=1}^4 \dd g^\epsilon_l \nn\\&&
\prod_{i=1,2} \la- \nb^{in}_{li}|{(g^{\epsilon}_1)}^\dagger g^\epsilon_2 |\nb^{in}_{ri}\ra^{2\gamma^\epsilon j^{in}_i}
\la- \nb^{out}_{li}|{(g^{\epsilon}_3)}^\dagger g^\epsilon_4 |\nb^{out}_{ri}\ra^{2\gamma^\epsilon j^{out}_i} \nn\\&&
\prod_{k=3,4} \la- \nb^{in}_{lk}|{(g^{\epsilon}_1)}^\dagger g^\epsilon_3 |\nb^{out}_{lk}\ra^{2\gamma^\epsilon j^{in}_k} 
\la- \nb^{in}_{rk}|{(g^{\epsilon}_2)}^\dagger g^\epsilon_4 |\nb^{out}_{rk}\ra^{2\gamma^\epsilon j^{in}_i} \delta(j^{in}_k, j^{out}_k)
\eea

We introduce the $\SO(3)$ group elements $G_{in}$, $G_l$, $G_r$ and $G_{out}$, related to the $SO(3)$ elements $G_i$ covered by the integration variables $g^+_i$ evaluated at the critical points by $G_{in} = G_1^\dagger G_2$, $G_{l} = G_1^\dagger G_3$, $G_r = G_3^\dagger G_4$ and $G_{out} = G_3^\dagger G_4$. The critical point equations then are:

\bea
-\nb^{in/out}_{li} =& G_{in/out} \nb^{in/out}_{ri} &\;\;\mbox{for} \, i = 1,2\nn\\
-\nb^{in}_{l/r\,i} =& G_{l/r} \nb^{out}_{l/r\,i} &\;\;\mbox{for} \, i = 3,4.
\eea

For non-degenerate boundary data the $\nb$ are pairwise linearly independent. Therefore these equations fix the $\SO(3)$ elements up to gauge, thus, as in the other cases, the second sector $\SU(2)\times\SU(2)$, $g^-$, has to agree with $g^+$ at the critical point up to gauge.

We now see that the geometric correlations induced by this intermediate amplitude are indeed in between the other two amplitudes. To see this consider the shape of each tetrahedron determined by the four areas given by the $j_i$ and the two exterior dihedral angles $\phi_{12}$ and $\phi_{34}$ given by $\cos(\phi_{ij}) = \nb_i \cdot \nb_j$ and $0 \leq \phi_{ij} < \pi$. By definition the areas of the left and right tetrahedra always coincide. In the case of the $A_4$ amplitude the asymptotic geometry is such that furthermore $\phi_{12}$ and $\phi_{34}$ coincide as well. We obtain the same geometry on the left and the right, which in turn eliminates the discontinuity at the equator, but leaves the $in$ and $out$ geometry completely uncorrelated. In the case of $A_0$, $\phi_{12}$ and $\phi_{34}$ in the left and right tetrahedron remain uncorrelated. Instead they now propagate from the $in$ state to the $out$ state. Furthermore now the four areas between the $in$ and $out$ state are forced to be the same, thus the entire (discontinuous) geometry from the $in$ state propagates to the $out$ state. For the $A_2$ amplitude we have exactly the intermediate situation. The areas $j^{in}_3$ and $j^{in}_4$ propagate to the areas $j^{out}_3$ and $j^{out}_4$, and the dihedral angles $\phi^{in}_{34}$ to $\phi^{out}_{34}$. They do however remain uncorrelated between the left and right tetrahedron on the $in$ and $out$ slice, and the geometry remains discontinuous. On the other hand the angles $\phi_{12}$ get matched between the left and right tetrahedron but do not propagate between $in$ and $out$.

We thus see that we have a trade off in the amplitudes between geometry propagating within the boundary and between boundaries. The only amplitude that both eliminates the discontinuity and induces correlations between the different boundaries is $A_3$.

In our discussion we have ignored the face amplitudes, as fixing these would require a separate discussion and separate assumptions. We refer the interested reader to the literature \cite{Bahr:2010bs} and references therein. To give the full asymptotics it would furthermore be necessary to systematically fix the phase of the intertwiners and understand the spin structure of the critical points.

\subsection{The consistency of the expansion}

As a sum over 2-complexes, the vertex expansion can be implemented as a Feynman expansion of an auxiliary field theory, called group field theory \cite{Ooguri:1992eb}. In the refinement limit it is more akin to a lattice approach. In both cases we are however lacking the physical expansion parameter that fixes the meaning of the expansion, in the former case the coupling constant associated to the vertices, in the latter case the lattice scale. Thus these analogies should be taken with a grain of salt. As mentioned above it has actually been suggested in \cite{Rovelli:2010qx} that both of these expansions can be considered to coincide for spin foam models, precisely due to the absence of a scale. However, we can still quite generically consider what sort of consistency conditions the amplitudes would need to satisfy for the low order calculations to be indicative of the behaviour of the full theory defined by either of these limiting procedures.

An example of such a consistency condition would be the invariance under Pachner moves used to construct lattice topological quantum field theories \cite{Fukuma:1993hy,Barrett:1993ab,Turaev1994,MR1273584}. In that case the 2-complex is taken to be the dual two skeleton of the triangulation of a PL-manifold \cite{nla.cat-vn1972092}. The condition then is that the observables on the boundary do not change under changes of the 2-complex that correspond to changing the triangulation of the manifold. A trivial refinement limit can then be taken that will depend on the PL-manifold and no longer on its triangulation. In particular observables can simply be calculated exactly using the simplest triangulations available.

The refinement limit considered in \cite{Rovelli:2010qx} envisions going to an ``infinite complete 2-complex'' and thus is of a significantly different nature. We can see this for example in the fact that as opposed to the Feynman expansion and the lattice approach the topology of the neighbourhood of a vertex is not restricted leading to infinitely many terms at the one vertex level already. We would also expect that in the case of gravity we will not be able to define the refinement limit exactly, but only approximately. Conversely that means that the consistency conditions should only be satisfied approximately, e.g. up to higher terms in a refinement parameter. In the case of perturbative renormalisation of QFT the statement is that the divergent parts of the higher order calculations resemble the lower order ones. In the case of non-perturbative renormalisation of lattice theories the hope is to take the refinement limit by approaching the critical point. In concrete lattice calculations the physical quantities are calculated for several values of the lattice spacing and then extrapolated to the continuum limit. Thus in this case, too, studying the behaviour under refinement is crucial for obtaining physical answers\footnote{E.g. \cite{Durr:2008zz} lists among the essential ingredients for a full and controlled study: ``V. Controlled extrapolations to the continuum limit, requiring that the calculations be performed at no less than three values of the lattice spacing, in order to guarantee that the scaling region is reached.''}. 


Dittrich and Bahr have emphasized the importance of studying the approximate behaviour of the theory and simpler models under such refinements, especially with respect to (restoring) their symmetries, and carried out a number of calculations in this direction for the classical \cite{Bahr:2009ku,Bahr:2009mc,Bahr:2009qc,Bahr:2010cq} and quantum case \cite{Bahr:2011uj}. In the case of general spin foam models Bahr has suggested a set of cylindrical consistency conditions \cite{Bahr:ta}. An analysis along these lines would be necessary to answer the question of the consistency of the approximation in the affirmative.

However, even without going to such lengths, and just using the operators defined above we can already formulate minimal consistency checks. Write $\tilde{A}_k$ for the sum of the $A_k$ operators with permuted edges and face factors taken into account. As the initial and final state on these operators have the same form, we can consider arbitrary combinations of these operators. These are the summands in $(\tilde{A}_0 + \tilde{A}_1 + \tilde{A}_2 + \tilde{A}_3 + \tilde{A}_4)^n$. Then, assuming the relative weight of terms depends only on the order of the vertex expansion, the sum over all spin foams with only these vertices is given by \be\label{eq:SF-sum}  A_{total} = \sum_n c_n (\tilde{A}_0 + \tilde{A}_1 + \tilde{A}_2 + \tilde{A}_3 + \tilde{A}_4)^n.\ee

For example, if $||\tilde{A}_0 + \tilde{A}_1 + \tilde{A}_2 + \tilde{A}_3 + \tilde{A}_4|| < 1$ and $c_n = 1$ we would obtain a close analogy to mass renormalisation in QFT $$A_{total} = \frac{1}{1 - (\tilde{A}_0 + \tilde{A}_1 + \tilde{A}_2 + \tilde{A}_3 + \tilde{A}_4)}.$$ If this term is structurally very different from the original $\tilde{A}_0 + \tilde{A}_1 + \tilde{A}_2 + \tilde{A}_3 + \tilde{A}_4$ term, the results of the first order calculation are meaningless for the behaviour of the full the theory.

Of course, the possibility of summing up these operators depends on the combinatorial factors in front of the sums, as well as possible face amplitudes that could easily render this sum strongly divergent. Furthermore it would require us to analyse the sum over spins in the theory for which no effective methods are available. Such an analysis is therefore unfortunately beyond the scope of this paper.

\subsection{The origin of factorisations}

We will now consider the factorisation of the amplitude $A_4$ discussed in \cite{Bianchi:2010zs}. Note that while $A_4$ factorises the entire amplitude, consisting of the 5 different terms as well as the identity operator, does not. The argument of \cite{Bianchi:2010zs} was that the amplitude $A_4$ should be understood as a Hawking-Hartle type no-boundary amplitude truncated to a single state:

$$A'_4 = \la\Psi_{HH}|\phi_{in}\tens\phi_{out}\ra,$$

for the state $\phi_{in}\tens\phi_{out}$ in the boundary Hilbert space. It was then further argued that as the classical constraint for the FRW universe factorises the state $\Psi_{HH}$ should be of the form $\Psi'_{HH}\tens \Psi'_{HH}$, and the amplitude should factorise:

$$A'_4 = \la\Psi'_{HH}|\phi_{in}\ra \la\Psi'_{HH}|\phi_{out}\ra.$$

However, this requires that the state encoded in the amplitude $A_4$ is a physical state in the sense that it is annihilated by the constraint. When viewed from the point of view of group averaging, which heuristically connects the Hamiltonian and the spin foam picture, first order terms should correspond to the matrix elements of the Hamiltonian, not eigenstates of the projector on its $0$ eigenvalue. Thus the factorisation of the constraint does not explain the factorisation of this contribution to the amplitude at first order.

Note that this factorisation property is very robust. The amplitude of every 2-complex that is connected only through one vertex, that is, that becomes disconnected upon removing one vertex, is equivalent to that of the disconnected 2-complex with an additional vertex. The amplitudes for such disconnected spin foams trivially factorise. Thus we can already see that an infinite number of terms in the graph and vertex expansion have this factorisation property.

Furthermore it should be pointed out that the factorisation is not a product of the asymptotic or semiclassical limit. It is exact for the full amplitude at this level of the vertex expansion. In particular this means that all correlation functions between the initial and the final state, including the graviton propagator, vanish.

From the boundary perspective we do not know the topology of spacetime, only of its boundary. That means that when we find the amplitude peaked it might be peaked on a solution to the classical equations on any spacetime with the boundary considered. A priori it is not clear that the topology of the 2-complexes dominating the peaked amplitude must correspond to the topology of the manifold on which the classical solution lives, on the other hand the very general factorisation property for 2-complexes connected at one vertex suggests that the amplitudes should correspond to classical solutions living on disconnected spacetimes. For the spin foam associated to $A_4$ the natural candidate is the union of two balls $B^4 \cup B^4$. In this case the Hartle-Hawking state will robustly factorise simply for topological reasons, no matter what approximations are performed on the disjoint spacetimes.

Another way to understand what type of space time topology the amplitude is seeing is by noting that the amplitudes we are considering are based on $\SU(2)$ BF amplitudes \cite{Barrett:2009as}. The suitably regularized $\SU(2)$ BF partition function is the integral over the flat connections on the 2-complex. In the case of triangulations this coincides with the integral over the flat connections on the manifold. Gluing the boundaries of the 2-complex corresponds to taking its trace, we can then see that constructing the operator $A_4$ from BF spin networks, instead of EPRL spin networks we have that \be tr(A^{BF}_4) = Z^{BF}(S^4). \ee That is, by identifying the components of the boundary $S^3 \cup S^3$ we obtain $S^4$ rather than $S^3 \times S^1$, further indicating that the space time topology the amplitude $A_4$ is $B^4 \cup B^4$.

We will see this interpretation further corroborated by the higher terms in the graph expansion we will discuss in the next section.


\section{The graph expansion}

In the graph expansion we are in a considerably better situation, as the refinement limit Hilbert space has been rigorously constructed \cite{Ashtekar:1993wf}. We know that the unrefined or truncated Hilbert space sits in the refined Hilbert space as a well defined subspace. In particular if the larger graph only contains one subgraph isomorphic to the unrefined graph this is the subspace given by setting the spin quantum numbers on all edges not in that subgraph to zero.

It had also been a long-standing problem how to find the homogeneous sector of this Hilbert space, for example to find what the right states corresponding to the Minkowski vacuum are. One way to attack this question is to attempt to embed the Hilbert space of loop quantum cosmology into the loop quantum gravity Hilbert space \cite{Bojowald:2001jz,Bojowald:1999eh}. This was successfully done by avoiding the no go theorem of \cite{Brunnemann:2007du} in \cite{Engle:2008in, Koslowski:2007nx}, and by changing the loop quantum cosmology Hilbert space in \cite{Brunnemann:2010qk,Fleischhack:2010zt}.

The crucial question for spin foam cosmology is then how the truncation to a small graph and the restriction to the homogeneous sector interact. In particular the states considered in the above embeddings generally have support on arbitrarily large graphs. Truncating to the small graph corresponds to setting quantum numbers to zero in the full theory. As these quantum numbers are the quantum numbers of geometry, the small graph truncation corresponds to having most areas and volumes evaluate to zero. This does not necessarily correspond to deep Planck scale behaviour, as the degrees of freedom that are excited can be arbitrarily large, but it is a very inhomogeneous distribution of quantum numbers throughout the graph.

\subsection{The 4-simplex graph.}

Keeping in the simplicial picture of this paper we will illustrate this issue by going to the complete graph on five vertices, $\Gamma^5$. This is the boundary spin network graph of a 4-simplex. In particular we will consider the analogue of the geometricity inducing amplitude $A_4$.

\begin{figure}[htp]
 \centering
 \includegraphics[scale=0.5]{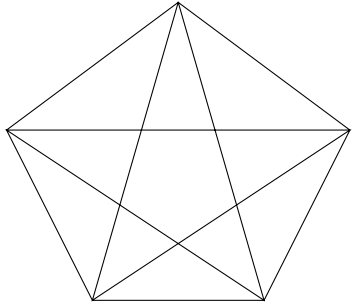}  
 \caption{$\Gamma^5$, the complete graph on five vertices. The vertices are labelled by $j,k = 1 \dots 5$, each edge is labelled by the end vertices and carries a spin $j_{kl}$, and each vertex carries an intertwiner $\iota_k$.}
 \label{fig-gamma5}
\end{figure}


The graph $\Gamma^5$ is dual to the smallest proper triangulation of the 3-sphere. The asymptotic geometry of the one vertex decomposable amplitude is given by the asymptotic analysis in \cite{Barrett:2009cj,Barrett:2009gg,Barrett:2009mw,Barrett:2009as}. It consists of two distinct flat 4-simplices with uncorrelated geometry. This reinforces the interpretation of the one vertex connected spin foam as having asymptotics dominated by bulk solutions of the classical theory living on the spacetime $B^4 \cup B^4$, rather than on the cosmological spacetime $S^3 \times [0,1]$.

To approximate the homogeneous sector we would usually choose a triangulation of the 3-sphere by equilateral tetrahedra. On the other hand we can embed the dipole graph into $\Gamma^5$ by setting the three edges connecting three vertices to zero, that is $$j_{34} = 0, \, j_{45} = 0, \, j_{35}=0.$$ This implies that $j_{13} = j_{23}$, $j_{14} = j_{24}$, and $j_{15} = j_{25}$ and we can embed the graphs considered above into $\Gamma^5$ by setting $j_{1i} = j_{i+1}$. We see that the subspace corresponding to the truncated graph is in fact orthogonal to the homogeneous sector.

In this scenario the geometry of the original graph inherited from the cylindrically consistent embedding into $\Gamma^5$ is that of a very degenerate and highly curved 4-simplex immersed degenerately into $\R^3 \subset \R^4$. In particular the integrated extrinsic curvature of the general $\Gamma^5$ geometric asymptotics is given by $$K^5 = \sum_{i,k \, i<k} \Theta_{ik}\, \gamma j_{ik},$$ where $\Theta$ is the exterior dihedral angle of the embedding of the 4-simplex in $\R^4$. This extrinsic curvature gives the phase of the asymptotics if the boundary is chosen to be the Regge state \cite{Barrett:2009cj}. Conceptually it corresponds to the integral of the action on the solution on which we are peaked. As the action in the interior is zero, this is the Regge analogue of the York-Gibbons-Hawking boundary term, that is, the integrated extrinsic curvature. Thus the appearance of this phase is considered the clearest confirmation that these amplitudes code a discrete gravity theory.

For the truncated graph the corresponding term will be $$K_4 = \pi\, \sum_{i=2}^5  \gamma j_{1i},$$ as every dihedral angle will go to $\pi$ or to $0$. As the sum of spins at each vertex needs to be integer the phase factor in this case will simply be a sign. We can see from the form of the integrated extrinsic curvature that the dynamics of the full theory treat this term as a highly degenerate configuration with no 4-volume. Remember that we considered only the non-propagating, geometricity inducing amplitude here. For more complex amplitudes we expect this problem to get worse as the inhomogeneities created by the cylindrical embedding will start propagating. Thus it appears clear that reducing to the homogeneous sector after truncating the boundary, while keeping the full dynamics in the sense of cylindrical consistent embeddings gives physically nonsensical results as the truncation will interact in essentially uncontrolled ways with the degrees of freedom we intend to capture.

Ignoring the simplicial picture we can still see this discrepancy. The extrinsic curvature of a homogeneous 3-sphere in $\R^4$ with equator of area $J = \sum_{i=2}^5 \gamma j_{1i}$ is given by $$K_{hom} = 2 \pi J,$$ that is, it is twice as large as the one seen in the phase of the amplitude.




\subsection{Truncated dynamics}

We have seen that the truncated state, when interpreted in the sense of cylindrical consistency, will look highly inhomogeneous to the dynamics of the theory. On the other hand there are natural notions of homogeneous states on the truncated theory e.g. in \cite{Rovelli:2008dx, Battisti:2009kp}. To take advantage of these it is necessary to consistently truncate the dynamics together with the state space. It is a priori unclear what this should mean for the sum over spin foams. A naive possibility would be to simply insist that the spin foam has a slicing that looks like the truncated boundary graph at all times, in which case only the identity operator and $A_5$ would contribute. This would also provide a restriction on the topology of the neighbourhood of the vertex necessary to render the one vertex expansion meaningful.

In the context of the Hamiltonian theory we can arrive at such a truncation by insisting that the matrix elements of the Hamiltonian for all larger graphs should vanish. It would be interesting to study the expansion of such a truncated Hamiltonian into spin foams, in particular whether it would be possible to express it in terms of the five basic amplitudes defined above.

Such Hamiltonians for truncated systems, in particular with an eye to the homogeneous sector, were discussed recently in \cite{Borja:2010gn,Borja:2010hn}. There the family of dipole graphs with $n$ legs connecting the two vertices is considered. The homogeneous states at each level are given by invariance under the truncated area preserving diffeomorphisms. The homogeneous subspaces are isomorphic for arbitrarily high $n$, but the isomorphism is not achieved by considering the cylindrically consistent embeddings of the smaller Hilbert spaces into the larger ones. For every particular truncation $n$ \cite{Borja:2010gn,Borja:2010hn} define a Hamiltonian consistent with the homogeneous sector. Interestingly in this case the restriction of the Hamiltonian to a subspace corresponding to a truncation $n'<n$ commutes with the action of the part of the symmetry group that leaves the subspace invariant. In this way this is an example of a system where symmetry reduction and cylindrical consistency commute, and as a consequence the large $n$ limit of the symmetry reduced sector is trivial.


Similarly in \cite{Battisti:2009kp} the discretised Hamiltonian of the truncated classical theory is considered. There is no reason to expect the dynamics induced by the Hamiltonian on a small triangulation to coincide with the dynamics induced by the Hamiltonian of a larger triangulation on the subspace corresponding to the further truncation to the small triangulation.

To derive the effective theory of the truncated state space from the dynamics of the full theory is already a difficult problem in the classical theory, see for example \cite{Buchert:2011yu}. Thus whether the ad hoc truncations would be indicative of the behaviour of the full theory would be hard detect. However, as in the case of the vertex expansion, one can study how the approximation behaves when switching on more degrees of freedom. For non-loop quantum cosmology this was already suggested and studied in \cite{Kuchar:1989tj}. To do so it will be necessary to understand how to embed the models of various complexity into each other in an appropriate sense.


\section{Conclusions}

We pointed out that even at the one vertex level arbitrarily complex amplitudes exist. Thus without further restrictions on the topology of the neighbourhood of the vertex the restriction to one vertex is too weak to be studied effectively. We then computed the five simplest terms of the first order in the vertex expansion, characterised by the minimal number of vertices and edges in the KKL network. We found an interesting trade off between geometricity and geometry propagation. Amplitudes that enforced a continuous geometry on the boundary did not propagate and vice versa. We discussed different coherence conditions necessary for the proposed expansions to work. We then showed that the factorisation property of the amplitude $A_4$ can be understood as a topological factorisation.

From our analysis it thus seems premature to claim that the lowest order terms show that there is a regime in which the theory reproduces cosmological spacetimes. Instead we find that the original calculation has to be reinterpreted as a calculation of the first topology changing spacetime. The topology of the type $B^4 \cup B^4$ is suggested by the general factorisation properties of one vertex connected spin foams, and corroborated by the calculation of the corresponding term in the next order of the graph expansion.

We then considered the meaning of the graph expansion in the sense of cylindrical consistency, and in particular its interaction with restricting to the homogeneous sector of the theory. We demonstrated the tension between the two by considering the next higher term in the graph expansion, the complete graph on five vertices, $\Gamma^5$. We concluded that in order to be able to consistently go to the homogeneous sector of the truncated theory, the dynamics of the theory need to be truncated, too. This is to not allow propagation to higher terms in the graph expansion, to which the truncated states look highly inhomogeneous. Such a truncation could help restrict the topology in the neighbourhood of the vertex, thus rendering the one-vertex expansion meaningful. The choice of the original paper could be seen as such a truncation. We then pointed out several proposals for truncated dynamics in Hamiltonian approaches to the dynamics. We pointed out that similar consistency conditions as for the vertex expansion have to be considered for the graph expansion as well. These can be seen as analogous to embeddings studied in the context of traditional quantum cosmology.

\section{Acknowledgements}

I would like to thank B. Bahr, M. Mart\'in-Benito, E. Bianchi, B. Dittrich, R. Dowdall, C. Rovelli, S. Steinhaus, S. Speziale and F. Vidotto for discussions and comments on a draft of this paper.

\bibliography{Foams}
\bibliographystyle{h-physrev4}

\end{document}